\documentclass[12pt,preprint]{aastex}

\shorttitle{Prediction of Solar Flares} \shortauthors{Qu et al.}

\begin{document}

\title{Prediction of Solar Flares from a Statistical
   Analysis of Events during Solar Cycle 23}

\author{Z. Q. Qu\altaffilmark{1}, C. L. Xu \altaffilmark{3}, X. L. Yan\altaffilmark{1,2},
Z.K.Xue\altaffilmark{1,2} and Z.N.Qu\altaffilmark{1,2}}

\altaffiltext{1}{Yunnan Astronomical Observatory, National
Astronomical Observatories, Chinese Academy of
              Sciences, Kunming, Yunnan 650011, P.R. China.}
\altaffiltext{2}{Graduate School of Chinese Academy of Sciences,
           Zhongguancun, Beijing, P.R. China.}
\altaffiltext{3}{Yunnan Normal University, Kunming, Yunnan 650092,
P.R.China.}
\begin{abstract}
Ways to give medium- and short-term predictions of solar flares are
proposed according to the statistical analysis of events during
solar cycle 23. On one hand, the time distribution of both C and M
class flares shows two main periods of 13.2 and 26.4 months in this
cycle by wavelet analysis. On the other hand, active regions of
specific magnetic configurations and their evolutions give high
productivity of C class flares but relatively low productivity of
energetic (M and X class) flares. Furthermore, by considering the
measurable kinetic features of active regions, i.e., the rotation of
the sunspots, some active regions of specified types are observed to
have high energetic flare productivity, above $66\%$. The
periodicity of the activity revealed can be used for medium-term C
and M class flare forecasting and the high productivity of active
regions forms the basis for short-term prediction of individual
energetic flares.
\end{abstract}


\section{Introduction}

     Flare prediction is an emergent but very difficult task for
solar physicists. Not only because the mechanisms involved in flare
occurrence are complicated (Tandberg-Hansen and Emislie, 1988), but
also the theories and observations are not very operationally
related. It is now well known that the short-term prediction of
individual flares should be done before or during the phase of
magnetic energy storage resulting in the energy release later.
Therefore, searching for sufficient or sufficient-like pre-flare
conditions is more helpful than searching for necessary ones in this
aspect. Statistical analysis is still a good tool to find such
conditions, and these sufficient or sufficient-like conditions must
correspond to the emergence of some determinant features of active
regions. In this paper, we try to find such conditions by the
statistics of events during solar cycle 23.

   Concerning the search for the pre-flare conditions, several ways
were proposed by investigators in the literature. For instance, the
most popular way is to find the correlations between sunspot
magnetic features and major flares(e.g. Greatrix 1963; Mayfield and
Lawrence, 1985; Atac 1987; Sammis et al., 2000; Ternullo et al.,
2006). Based on this kind of correlation and solar cycle data,
Qahwaji and Colak (2007) established an automatic short-term
prediction system with an ability of machine-learning. On the other
hand, some statistical correlation between flare productivity and
more detailed magnetic features, like maximum horizontal gradient of
photospheric vector magnetic fields, length of neutral lines, and
number of singular points, as well as magnetic shear, has been done
by Cui and Wang (2007), and Cui et al.(2007). They found the
correlation can be fitted by Boltzmann sigmoid functions and
proposed that this relationship can be used for modeling flare
forecasting.

At the present stage, we are far from forecasting all the flares or
even only significant flares individually. However, it is very
useful to give medium- and short-term predictions of some individual
energetic(M and X class) flares. This is the main purpose of this
paper. Gallagher, Moon and Wang(2002) based on the flare
productivity of active regions, proposed some kind of flare
forecasting, giving even 10 variables including McIntosh
classification, sunspot group area, the number of sunspots, the
average hardness indexes, etc. in their model. Like Jakimiec and
Bartkowiak (1994), other similar features were included(Zhu and
Wang, 2003). However, the short-term prediction of a great number of
individual flares is far from satisfactory (Zhu and Wang, 2003).
Therefore, searching the determinant variables which promote the
prediction is crucial. Another way was proposed by Wheatland(2001),
that the rate of flare production in single active regions from 1981
to 1999 can be obtained by studying the waiting-time
distribution(WTD) for individual regions and the relation reflecting
WTD, say, described by Poisson process can be used for the
prediction. Evidently these two methods are complementary and can be
used for further investigations in this field.

\section{Periodicity of Activity and High Flare Productivity Found during Solar Cycle 23}
   Solar cycle 23 provides some clues. Gaining some
characteristics of this cycle can help us make the predictions for
the next cycle. All of the data for the events used for our analysis
stem from the website http://www.solarmonitor.org/. When counting
flares, we should be careful to get rid of recounted flares recorded
on the site in neighboring days. The time span covers the whole of
solar cycle 23, i.e., from Feb. 1996 to Jan.2008. The total number
of flares recorded by the website is 21,620, while 19,577 flares
were classified as C class, 1,887 as M class and 156 as X class
according to soft X ray flux classification. There were 11,138
(10,516 C class, 587 M class and 35 X class) flares not related to
numerated active regions. They are also included for the statistical
analysis. In order to have an overview of these recorded flares,
Fig.1 outlines the number distribution of the flares of different
soft X ray flux. One can see that the M1 class of flares stands out
as an exception and this may become a feature of the activity of the
cycle . We use the exponential function
\begin{equation}
       N(x)=a+be^{-c log_{10}x}
\end{equation}
to describe the number distribution (solid line), where $x$ means
the soft X ray flux irradiated by flares. The coefficients ($a=76,
b=2.0\times 10^{-5}, c=4.07$) can become other characteristic
quantities describing the activity of this cycle. In the fitting, we
do not include either the faintest flares (C1.0) recorded, because
the number of these flares are likely underestimated,  or the
highest energetic flares above X10.0 because they are very rare.

\begin{figure}
\centering
\includegraphics[width=7.4cm,height=5cm]{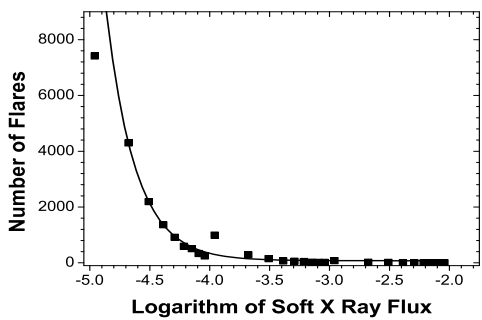}
\caption{Number distribution of flares to logarithm of soft X ray
flux irradiated by flares in unit of erg/cm$^{2}$s. The solid line
is the fitting by Eq.(1) as stated in the text.} \label{}
\end{figure}

In the upper panel of Fig.2, the number evolution of events with
solar flare classification is depicted. One can see that the pattern
takes on periodicity except for X class flares, whose number
evolution is not plotted individually. This stimulates us to use the
wavelet analysis with the Marlet function as the wavelet function to
find the periods. In the lower panel, the corresponding power
spectra are shown. It is obvious that all the curves have two
prominent peaks at the 13.2 and 26.4 months for both C class and  M
class flare occurrence, and finally for all the recorded flares.
Furthermore, the phases are also very close to each other. Such
periodicity can be used for medium-term prediction of the next
cycle, although the periods cannot be expected to be same. Finally,
we may assign these periods as features of this cycle.

\begin{figure}
\centering
\includegraphics[width=9.0cm]{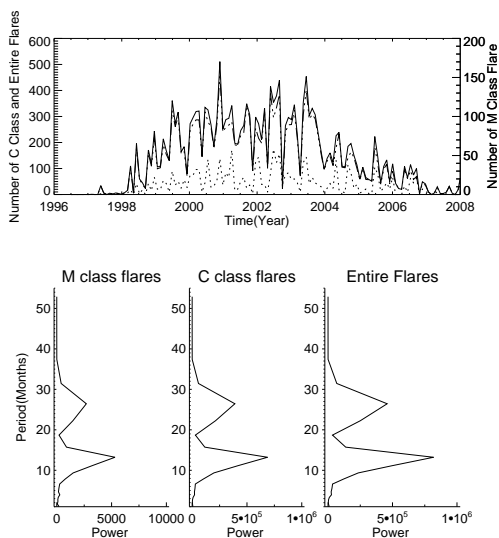}
\caption{The number evolution of C class flares(dotted line), M
class flares(dashed line) and all classes of flare number(solid
line) above C1.0 class during solar cycle 23(upper panel), and the
power spectra of flare occurrence obtained by wavelet analysis
(lower panel).} \label{}
\end{figure}

   It is generally accepted that special magnetic configurations
and their evolution induce flare eruptions. At the present stage,
the three dimensional classification is not at hand for all the
numerated active regions and the two dimensional Hale and McIntosh
classifications are more frequently used. The lower panel of Fig.3
presents the numbers of the active regions belonging to particular
magnetic configurations (Hale classification) without any
configuration variation within one day. A similar diagram also
appeared in the work by Ternullo et al.(2006) but only into June
2003. In the following, the flare productivity is defined as the
ratio of the number of active regions hosting flares of specific
class to the total number of active regions of the same specified
type. The configuration $\beta$ is dominant in this cycle. However,
the flare production does not favor of this configuration or another
simple one, $\alpha$. High flare productivity occurs in other
configurations, i.e. $\beta\delta$ ($64\%$, 25 regions hosting
flares in 39 regions owned this configuration), $\beta\gamma$
(53$\%$, 547 in 1033), $\beta\gamma\delta$ ($75\%$, 300 in 402),
$\gamma$ (100$\%$, 1 in 1) and $\gamma\delta$ (83$\%$, 5 in 6)
(upper panel). For these regions containing the listed
configurations, the M class flare productivity reached $28\%$(11 in
39) in the active regions containing configurations of $\beta\delta$
producing 18 M class flares; $10\%$( 102 in 1033) in regions of
$\beta\gamma$ hosting 139 M class flares; 34$\%$(138 in 402) in
regions of $\beta\gamma\delta$ inducing 256 M flares, and for rare
configurations, 100$\%$(1 in 1) productivity in the region of
$\gamma$ and 50$\%$(3 in 6) in regions of $\gamma\delta$. Finally
for X class flare production, the highest efficiency took place in
active regions containing configurations of $\beta\gamma\delta$
hosting 46 X class flares by 38 of 402 regions, with productivity of
only 9$\%$; and 8$\%$ in those of $\beta\delta$ producing 4 X class
flares. This statistics reveals the conclusion that energetic flares
are not only relatively favored regions classified
$\beta\gamma\delta$(Sammis et al., 2000) but also more
configurations relatively. But the absolute productivity is not so
high for those regions hosting more than 10 energetic flares. A
similar conclusion has been drawn by Ternullo et al. (2006) but with
a shorter time span in the cycle.

\begin{figure}
\centering
\includegraphics[width=9cm,height=7cm]{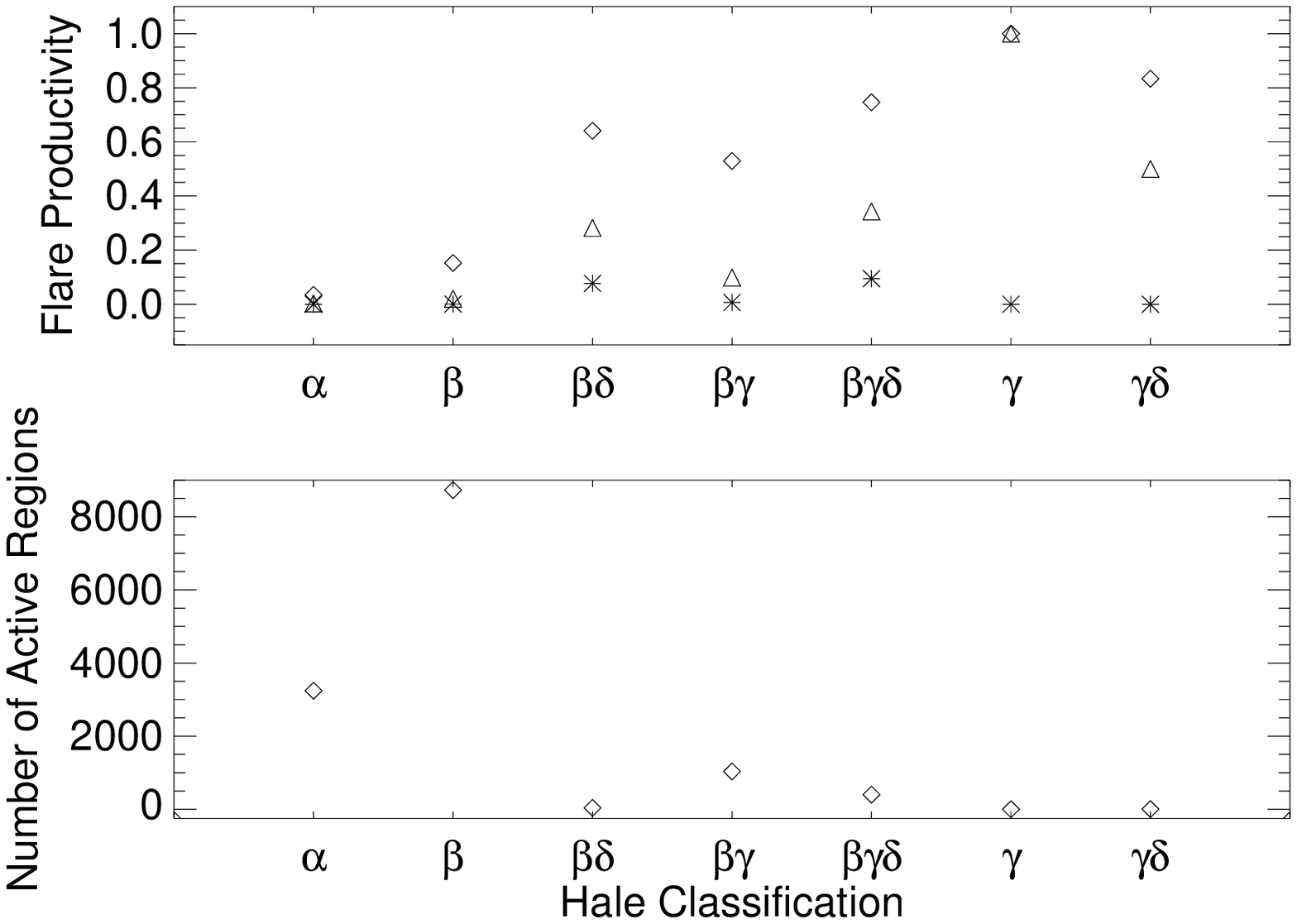}
\caption{The flare productivity of different magnetic configurations
(upper panel) and the numbers of these active regions of
corresponding types (lower panel). The symbol $'\diamondsuit'$
represents the productivity for all classes of flare while
$'\triangle'$ for M class flares and $'*'$ for X class.} \label{}
\end{figure}
   On the other hand, detectable variations of configuration
can reveal another way to get the flare production efficiency.

  In the top panel of Fig.4, we plot the productivity of 13 magnetic
configuration evolution patterns among 31 patterns during the cycle.
These patterns are selected by the criterion that the number of
active regions that underwent one of these evolutions is more than
9. In the bottom panel is illustrated the number of corresponding
active regions. It should be noted that some active regions could
experience several evolution patterns. The symbol $'\diamondsuit'$
indicates the productivity including all flares above class C1.0,
while the symbol $'\triangle'$ stands for the productivity for M
class flares and symbol $'*'$ for X class flares. The highest
productivity (75$\%$) of flares of all classes involves the
evolution from $\beta\gamma$ to $\alpha$, though the most frequent
configuration change pattern was from $\alpha$ to $\beta$. In fact,
there are 6 important evolution patterns that show very high
productivity (above 55$\%$). They are $\beta\delta\rightarrow
\beta$, $\beta\gamma\rightarrow \alpha$,
$\beta\gamma\delta\rightarrow \beta$, $\beta\gamma \rightarrow
\beta$, $\beta\gamma \rightarrow \beta\gamma\delta$ and
$\beta\gamma\delta \rightarrow \beta\gamma$. On the other hand, the
M or X class flares were not highly favored by any those 13 kinds of
active regions. The highest M class flare productivity ($33\%$)
occurred in the regions experiencing evolution from $\beta\gamma$ to
$\alpha$, and only 10 M class flares were produced in 4 regions. It
was also this evolution pattern which contained the highest X class
flare rate of production among the 13 evolution patterns. 2 among 12
of such regions produced 5 X class flares with a productivity of
17$\%$. The relatively high M class productivity occurs in the
evolution patterns from $\beta\delta\gamma$ configuration to other
configurations. For instance, 33$\%$(1 in 3) in the regions of
configurations from $\beta\delta\gamma$ to $\alpha$, 30$\%$(16 in
53) to $\beta$, 33$\%$(3 in 9) to $\beta\delta$, 50$\%$(1 in 2) to
$\beta\gamma$. And finally 30$\%$(36 in 118) from
$\beta\delta\gamma$ to $\beta\gamma$.

For very rare evolution patterns, like that from $\gamma$ to
$\beta\gamma\delta$, 100$\%$ M class productivity has been reached
but only one active region once experienced such evolution. However,
the very low X class flare production rate was evidenced for the
most frequently evolution patterns. The rate of production is only
12.5$\%$ (1 in 8 regions) obtained by the evolution from
$\beta\delta$ to $\beta\gamma\delta$ and 11$\%$(1 in 9 regions) from
$\beta\delta$ to $\beta\gamma$.

\begin{figure}
\centering
\includegraphics[width=9cm,height=8cm]{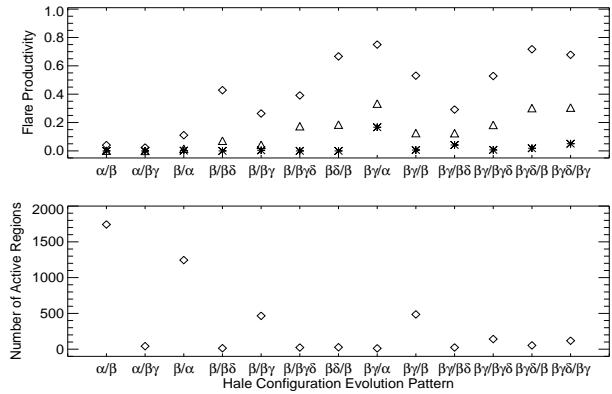}
\caption{The flare productivity of different magnetic configuration
evolution patterns (upper panel) and the numbers of these active
regions (lower panel). The symbol $'\diamondsuit'$ represents the
productivity for all classes of flare while $'\triangle'$ for M
class flares and $'*'$ for X class.} \label{}
\end{figure}

\section{One Way to Give Short-term Prediction of Solar M and X Class Flares}

   In the above analysis, we see the statistical correlation of the
Hale classification and its evolution to flare productivity obtained
from solar cycle 23. Evidently, some specific configurations and
 evolutions of active regions have very high flare production
efficiencies. However, for M and especially X class flares (which
are significant for the prediction), we did not see enough high
productivity (say, above 50$\%$), with enough flares, by observing
the magnetic features of active regions. Naturally, one always
expects to forecast more individual energetic flares. Therefore, to
get a valuable and accurate warning, one should search more
determinant features for those active regions hosting energetic
flares.

   One way is to consider the real time measurable kinetic features of
sunspots among which two features, to our knowledge, are good
choices, i.e., the emergence of new flux (Li et al., 2007) and
rotation of the sunspots (Yan et al., 2007) . Here, we focus on the
latter, and try to analyze the energetic flare productivity of
active regions having sunspot rotations of different types, along
with their magnetic configurations.

   During solar cycle 23, 182 rotating sunspots
from December 1996 to December 2007 were found (Yan et al., 2008a).
Among them, 35 X class flares were related to 17 active regions with
their sunspots rotated before flare occurrence. The rotation causes
the energy accumulation of the flare (see Yan et al., 2007). By
tracing the rotation and the following flare occurrence, we give the
statistical analysis of the correlation between flares and
rotations. In our recent research (Yan et al., 2008b), sunspots with
rotations are classified into several types related to their
magnetic configurations and the productivity was listed.

   Figure 5 illustrates the time difference between the recognizable
start time of rotation and the onset time of X class flares. It
should be pointed out that the start time of sunspot rotation
depends on the data sources (Yan et al., 2008b), typically the error
is about 2 hours. One can see from the figure that the smallest time
difference is a little longer than 2 hours while the greatest
difference is less than 150 hours (a little longer than 6 days),
which happened for type Vb. On the other hand, some of these regions
produced more than one flare, and the time difference between
successive flares within the same region can also be seen from the
figure. The definition of these rotation types related to the
magnetic configuration can be found in Yan et al.(2008b).

  From the statistics, the energetic M and X class flare productivity
can reach 100$\%$ in active regions of special types. For instance,
in those active regions with rotating sunspots and magnetic
configurations of type Vb (NOAA10930, NOAA09236, the only two
regions found belonging to the type in the cycle), there were 5 X
class flares occurring in the two regions. For M class flares,
100$\%$ production efficiency was met by types IVc, IVd and Vb.
 5 M class flares occurred in NOAA09070 of type IVc, 1 M
class flare happened in NOAA10464 of type IVd, 9 M flares took place
in NOAA10930 and NOAA09236 of type Vb. Therefore, active regions of
type Vb are very productive with respect to energetic flares! This
type of active region was defined as follow: in these regions there
is one sunspot that not only spins but also rotates around another
sunspot with opposite polarity(Yan et al.,2008b).

  Other types of rotation patterns which exist in the active
regions can also induce high production rates of M and X class
flares. For instance, for M class flare, types IIa, IIIa, IIIb, and
VIb can have more than 66$\%$ productivity. In detail, 6 in 8 active
regions of type IIa manufactured 42 M class flares, with
productivity of $75\%$; 5 in 6 active regions of type IIIa gave
birth to 31 M class flares and 5 in 6 regions of type IIIb produced
42 M class flares with the same productivity of 83$\%$; 2 in 3
regions of type VIb induced 8 M class flares, with productivity of
$67\%$.

\begin{figure}
\centering
\includegraphics[width=7.0cm, height=4.0cm]{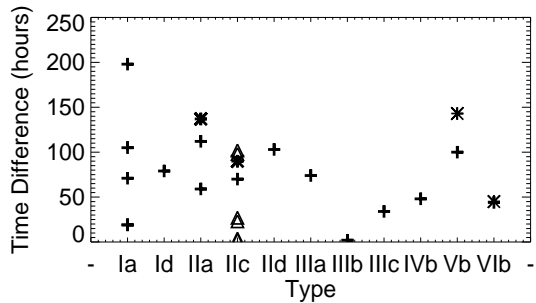}
\caption{Time difference between the detectable beginning time of
sunspot rotation and the onset of X class flares. In the diagram,
for the same type of rotation in the vertical direction, different
symbols indicate different active regions.} \label{}
\end{figure}

    Therefore, we conclude that, if the measurable
kinetic features of sunspots in active regions are taken into
account, one may obtain a very high reliability for energetic flare
prediction.

\section{Conclusions}

   Evidently, the most difficult mission for solar physicists is to
forecast flares far from sunspot-inhabited regions. However, in
active regions, one can find some clues to issue even short-term
predictions of individual flares as we provided in the above
section. At the present stage, researchers cannot give exact
short-term predictions for all individual solar flares. Therefore,
one way is to issue predictions with high reliability.

 In this paper, we find C and M class flare occurrence periods of 13.2
months occurring twice during solar cycle 23 and we believe that
such activity periodicity may exist in the past cycles or the coming
cycles. This can be used for medium-term forecasting. Additionally,
some special magnetic configurations and their evolution patterns
can also help us present short-term prediction. They are listed as
follows:

1) Magnetic configurations. Usually, the more complicated are the
configurations of the active regions, the higher the productivity
they have. For all kinds of  flare classes, configuration
$\beta\gamma\delta$ has the highest rate of production, followed by
$\beta\gamma$;

2) Magnetic configuration evolution patterns. The flares are favored
by evolution patterns from configuration $\beta\gamma\delta$ to
other ones.

  However, as pointed out above,  one cannot get high reliability for
predicting energetic flares by only observing the two dimensional
magnetic configurations and their evolutions. But we also showed
that it is possible to make short-term predictions with high
reliability of more individual energetic flares if the real time
measurable kinetic features of active regions, like sunspot
rotation, are taken into account. Especially, we listed the sunspot
rotation patterns of active regions related to magnetic
configurations that showed high productivity of energetic flares as
follows:

1) High X class productivity. Active regions having sunspot rotation
type Vb(100$\%$);

2) High M class productivity. Active regions having sunspot rotation
types IVc, IVd, and Vb have 100$\%$ productivity and active regions
of types IIa, IIIa, IIIb, and VIb have production rates above
66$\%$.

   The features of active regions having 100$\%$ M or X class flare
productivity are candidates for the sufficient conditions we are
searching for. Other features with high productivity, e.g., above
66$\%$, are also very useful. Of course, more real time measurable
features can be included, for instance, the emergence of new flux
into the active regions (Li et al., 2007). Therefore, it is
reasonable to expect that more and more energetic flares can be
predicted. On the other hand, one can also consider the WTD for
these specific individual regions, like the work done by
Wheatland(2001). It is evident that combining these two methods will
promote short-term prediction for more and more individual flares.

 Finally, it should be pointed out that the productivity of the above specified
types of active regions may experience fluctuations from one cycle
to another. However, our work here provides some clues or methods to
promote medium- and short-term flare prediction.

\acknowledgments {{\bf Acknowledgments} \footnotesize This work is
supported by the National Science Foundation of China (NSFC) under
the grant number 10673031 and 40636031, National Basic Research
Program of China 973 under the grant number G2006CB806301. The
authors are grateful to the BBSO, TRACE, SOHO/MDI and HINODE
consortia etc. for their data.}

\end{document}